# On correlation lengths of thermal electromagnetic fields in equilibrium and out of equilibrium conditions

Illarion Dorofeyev[1]

[1]*Institute for Physics of Microstructures Russian Academy of Sciences, GSP-105, Nizhny Novgorod, 603950 , Russia*
*Corresponding author: Illarion1955@mail.ru*

Spatial coherence of thermal fields in far- and near-field zones generated by heated half-space into vacuum is studied at essentially different thermodynamical conditions. It is shown that correlation lengths of fields in any field zone are different in equilibrium and out of equilibrium systems. In wide range of distances from sample surface correlation functions are should be calculated using a total sum of evanescent and propagating contributions due to their mutual compensation at some conditions because of anticorrelations. It is demonstrated that correlation lengths as calculated with a proposed formula are in agreement with a behavior of correlation functions of thermal fields in spectral range of surface excitations

## 1.Introduction

The study of the time and spatial coherence properties of spontaneous fields is an important part of modern science. Correlation properties of free thermal electromagnetic fields in equilibrium with surrounding matter are described in detail; see for example [1-4]. Correlation tensors of the fields at any distances from bodies with arbitrary geometric shapes can be theoretically calculated both in equilibrium and in out of equilibrium when heated body radiates into a cold surrounding [5-8]. It should be emphasized that in accordance with the theoretic analysis thermal fields are formally divided into the propagating and evanescent fields with different spectral and correlation properties, see for numerous examples [9-11]. We recall that the correlation radius of propagating waves of the black body field is on the order of the Wien wavelength in case of non-equal time correlation functions, and is on the order of a wavelength of interest in case of equal time correlation functions. The spatial correlation scale of fields nearby heated bodies depends on electrodynamic properties of materials and other parameters of the problem under study. An analysis of properties of thermally stimulated fields in a vicinity of interfaces is extremely important for a description of dynamics of near surface processes. There has been a vast amount of scientific activity in the study of the coherence properties of thermal fluctuations at frequencies of the surface excitations and waveguide modes [12-16]. The prominent result here is the increase of the spatial correlation at the surface eigenfrequencies. The increase can reach tens of wavelengths nearby an interface which is substantially larger than the correlation scale of the equilibrium radiation in free space [12]. Furthermore, it was shown in [11, 15] that the length of correlation is determined by the dispersion characteristic for surface polaritons. The physical grounding is that the collective coherent excitation of surface charges or surface oscillations of the lattice at the interface becomes as the source of fields at corresponding eigenfrequencies transferring the spatial coherence to the fluctuating electromagnetic fields. Obviously, a study the coherence of spontaneous fields in the frequency range of resonance states or eigenmodes is especially important, because properties of eigenmodes are completely determined by whole set of characteristics of the system.

Our paper addresses the following question: how different the coherence lengths of thermal electromagnetic fields in equilibrium and out of equilibrium systems. In sections 2 we give general formulas for traces of spatial coherence tensors of the fluctuating electromagnetic fields generated by a half-space into cold space, and radiated by a half-space into surroundings kept with the same temperature. Numerical results for coherence tensors of thermal fields and for the correlation length within thermodynamically different systems and relating discussions are provided in section 3. Concluding remarks are given in section 4.

## 2. Spatial correlations of thermal fields

We consider a system of heated half-space radiating into a vacuum in different thermodynamic conditions. One condition corresponds to the complete thermodynamic equilibrium and another condition corresponds to the system in which the temperature of the half-space is much larger than the temperature of the ambient bodies at infinity. In both cases the general formulas for the cross-spectral tensors of thermal fields at any distance from the half space with flat boundary can be found in [6, 9-11]. All spatial correlations are considered in vacuum over a heated half-space between points $\vec{r}_1 = (0,0,h)$ and

$\vec{r}_2 = (L\cos\Phi, L\sin\Phi, h+H)$, where $h$ and $h+H$ are the distances along the normal to the boundary of a half-space, $L$ is the separation of the selected points along the interface. The angle $\Phi$ will be irrelevant after taking a trace over the cross-spectral matrices $w_{ij}(\vec{r}_1, \vec{r}_2; \omega)$, $(i,j=x,y,z)$. It is remarkably that any statistical characteristics of thermal fields in the problems of our consideration can be divided into two separate parts corresponding to the evanescent and propagating waves. The same is valid for traces of the cross-spectral matrices. Thus in case when the heated half-space radiates into cold surroundings the trace of the cross-spectral matrix for components of the electric field can be represented as follows

$$W^{neq}(L,H;\omega) \equiv Trace\{w_{ij}^{neq}(L,H;\omega)\} = f^{Pr}(L,H;\omega) + f^{Ev}(L,H;\omega), \quad (1)$$

where

$$f^{Pr}(L,H) = \pi u_{0\omega} \int_0^{\pi/2} d\theta \sin\theta \cos(k_0 H \cos\theta) \times [(1-|r^P|^2)/2 + (1-|r^S|^2)/2] J_0(k_0 L \sin\theta), \quad (2)$$

$$f^{Ev}(L,H) = \pi u_{0\omega} \int_0^\infty dy \exp[-k_0 y(2h+H)] \times [(2y^2+1)\text{Im}\{r^P\} + \text{Im}\{r^S\}] J_0(k_0 L \sqrt{1+y^2}), \quad (3)$$

where $r^{P,S}$ are the Fresnel coefficients, $u_{0\omega} = \Theta\omega^2/\pi^2 c^3$ is the spectral power density of the black-body radiation, $\Theta(\omega,T) = (\hbar\omega/2)cth(\hbar\omega/2k_B T)$, $J$ is the Bessel function, $k_0 = \omega/c$.

In equilibrium case the trace is expressed by the following formula

$$W^{eq}(L,H;\omega) \equiv Trace\{w_{ij}^{eq}(L,H;\omega)\} = g^{Free}(L,H;\omega) + g^{Pr}(L,H;\omega) + g^{Ev}(L,H;\omega), \quad (4)$$

where

$$g^{Free}(L,H) = 2\pi u_{0\omega} \int_0^{\pi/2} d\theta \sin\theta \cos(k_0 H \cos\theta) \times J_0(k_0 L \sin\theta), \quad (5)$$

$$g^{Pr}(L,H) = \pi u_{0\omega} \int_0^{\pi/2} d\theta \sin\theta J_0(k_0 L \sin\theta) \times \text{Re}\{(r^S - r^P \cos 2\theta) \exp[ik_0(2h+H)\cos\theta]\}, \quad (6)$$

and $g^{Ev} = f^{Ev}$ form Eq.(3).

It should be noted that the evanescent contributions of the traces $f^{Ev}$ and $g^{Ev}$ are identical in equilibrium and in nonequilibrium problems.

Using the formulas for traces we numerically studied the correlation length of thermally stimulated fields generated by the material half space whose dielectric functions are given by the Drude or by the oscillatory models.

## 3. The correlation length of thermal fields

It should be emphasized that no stricly defined notion of the correlation length following from some basic physical principles. The experimentally meausred value is the correlation function. The correlation length itself is the measure of the "essentiality" of correlations of some process. It may be measured, for example from 50 per cent up to 1 per cent level.

For instance, from dimensionality reasons the correlation length along some $X$ direction can be defined

$$\ell_{(z)}^X = \frac{2\int_0^\infty dX |Tr\{w_{ij}(X)\}|^z}{|Tr\{w_{ij}(X=0)\}|^z}, \quad (7)$$

where $z$ is any real numbers, in the simplest case they are natural numbers $z = n = 1, 2, 3, \ldots$.

Here we take the often-used expression for correlation length along, for example $L$ direction as follows

$$\ell_{corr}^L = \frac{2\int_0^\infty dL |Tr\{w_{ij}(L)\}|^2}{|Tr\{w_{ij}(L=0)\}|^2}, \quad (8)$$

and similar expression for $H$ direction.

### 3.1. The correlation length in free space in equilibrium and out of equilibrium cases

In equilibrium in free space, or sufficiently far from any boundaries of a cavity the thermal electromagnetic field is the black-body radiation. In this case all chracteristics are known and we obtain a numerical value of the correlation length using different approaches. For example, the normalized cross-spectral tensor in Eq. (3.15) from [3] is as follows

$$Tr\{w_{ij}(\vec{R})\} = 3\sin\tilde{R}/2\pi\tilde{R}, \quad (9)$$

where $\tilde{R}$ represents the separation in units of the wavelength $\lambda$ of two arbitrary spatial points. Taking into account Eqs.(8) and (9) we have the corrrellation length $\ell_{corr}^R = \lambda/2$ along the arbitrary direction $R$.

From other side, in equilibrium and far ($k_0 h \gg 1$) from the heated half-space we obtain from Eq.(4) that $W^{eq}(L,H) \simeq g^{Free}(L,H)$, where $g^{Free}(L,H)$ is given by Eq.(5). In this case using Eq.(8) we have the identical correlation lengths $\ell_{corr}^H = \ell_{corr}^L = \lambda/2$ both along the direction $H$ (L=0) and $L$ (H=0). We note that the correspondence between these distances is $\tilde{L}^2 + \tilde{H}^2 = \tilde{R}^2$.

Thus, we recall the well known results that in equilibrium in free space the corrrellation length is in order of the wavelength of the black body radiation. Besides, other characteristics of the black-body field are independent on optical properties of surrounding bodies.

In its turn in the out of equilibrium case at $k_0 h \gg 1$ we have from Eq.(1) that $W^{neq}(L,H) \simeq f^{Pr}(L,H)$, where $f^{Pr}(L,H)$ is given by Eq.(2). The spatial correlations become the frequency dependent characteristics. For numerical calulations we have chosen the following parameters of matter: the plasma frequency $\omega_p = 122000\, cm^{-1}$ and damping $\nu = 0.01\omega_P$ for the Drude model, the frequencies of transversal and longitudinal phonons $\omega_T = 793\, cm^{-1}$, $\omega_L = 969\, cm^{-1}$, the damping $\gamma = 4.76\, cm^{-1}$ and $\varepsilon_\infty = 6.7$ for the oscillatory model corresponding SiC.

Figure 1 exemplifies the normalized correlation length $\ell_{corr}^{H,L}/\lambda$ in units of the wavelength $\lambda$ of thermal fields in free space versus a frequency in case when a heated half-space radiates into cold surroundings. Fig.1a)- for metal, Fig.1b) – for SiC. Upper curves in both figures correspond to the correlation length along of the $H$ direction, and low curves – along of the $L$ direction.

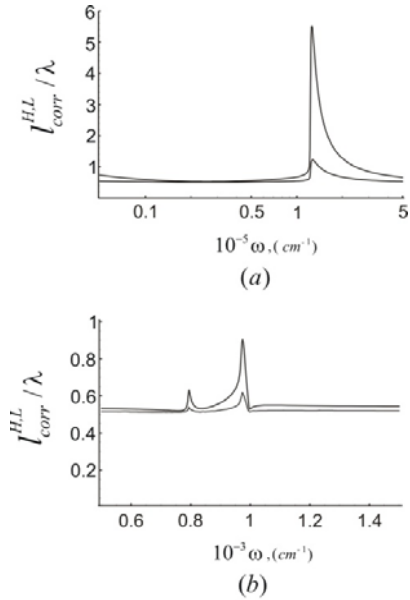

Fig. 1. Normalized correlation length $\ell_{corr}^{H,L}/\lambda$ of electrical part of thermal fields in free space versus a frequency in case when a heated half-space radiates into cold surroundings. Fig.1a)- for metal, Fig.1b) – for SiC.

It is clearly seen from the figure that the correlation lengths are increased at the eigen frequencies of the systems. In case of a metal the correlation length peaks at the plasma frequency $\omega_p = 122000\, cm^{-1}$, and in case of SiC – at the frequencies of the transversal $\omega_T = 793\, cm^{-1}$ and longitudinal $\omega_L = 969\, cm^{-1}$ optical phonons. It should be emphasized that the correlation length of thermal fields within a nonequilibrium system is larger than the correlation length within a system in equilibrium.

### 3.2. The correlation length nearby surface sample in equilibrium and out of equilibrium cases

Properties of fluctuating electromagnetic fields are especially important in the nearest vicinity of interfaces both in equilibrium and out of equilibrium problems. A possible way to descibe the fields just at interfaces was proposed in [17]. As we have already mentioned the spectral and correlation properties of thermal fields can be computed based on the classical electrodynamics and using the fluctuation dissipation theorem [5-11]. This approach is valid at distances from interfaces which are much larger than the interatomic distances, at least. Different contributions in traces appear additively and independently of each other in Eqs.(1) and (4). Figure 2 repersents normalized values $\tilde{g}^{Free}(L) = g^{Free}(L,0)/\pi u_{0\omega}$ - (a), $\tilde{g}^{Pr}(L) = g^{Pr}(L,0)/\pi u_{0\omega}$ - (b) and
$\tilde{g}^{Ev}(L) = g^{Ev}(L,0)/\pi u_{0\omega}$ - (c) of different contributions to the total trace as the functions of normalazed lateral separations $L/\lambda$ in accordance with Eqs.(4)-(6). The functions $\tilde{g}^{Pr}(L)$ and $\tilde{g}^{Ev}(L)$ are computed at the fixed distance $h/\lambda = 0.3$.

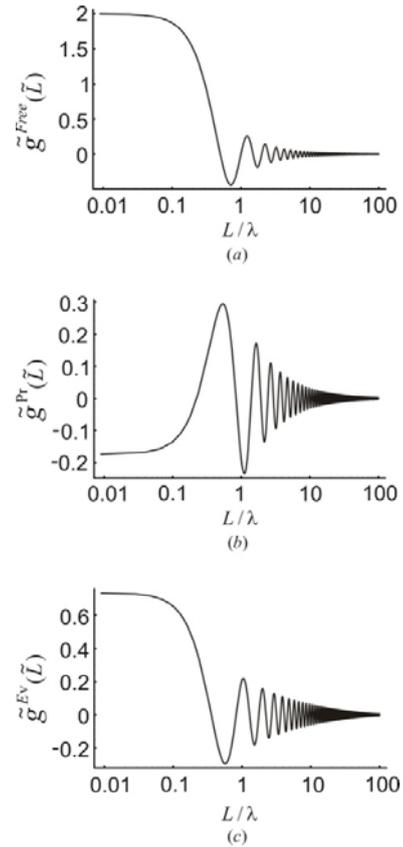

Fig.2. Normalized values $\tilde{g}^{Free}(L)$ - (a), $\tilde{g}^{Pr}(L)$ - (b) and $\tilde{g}^{Ev}(L)$ - (c) of different contributions to the total trace as the functions of normalized lateral separations $L/\lambda$ in accordance with Eqs.(4)-(6).

It is clear that in nature only the sum of these contribution is relevant at any distance from a sample surface. We note here that in case of spectral power density it is reasonably valid to neglect by the propagating or evanescent contributions in the near- or far-field zones. But, it should be careful with neglections in studying of correlation functions at different distances from an interface. The cross-spectral tensors have alternating signs corresponding to the correlations or anticorrelations. The first term in Eq.(4) is the distance independent term relevant to all distances including the near-field zone. Moreover, the second term in Eq.(4) corresponding to Fig.2b) starts with anticorrelations. This term is comparable with other two at distances $h/\lambda > 0.01$ and effectively compensates them due to the anticorrelations. That is why we must taking into account all contributions to descibe correlations correctly.

In it's turn, Fig.3 demonstrates normalised values $\tilde{f}^{Pr}(L) = f^{Pr}(L,0)/\pi u_{0\omega}$ - (a) and $\tilde{f}^{Ev}(L) = f^{Ev}(L,0)/\pi u_{0\omega}$ - (b) in case of the system out of equilibrium with use of Eqs.(1)-(3). The function $\tilde{f}^{Ev}(L)$ is computed at the fixed distance $h/\lambda = 0.3$.

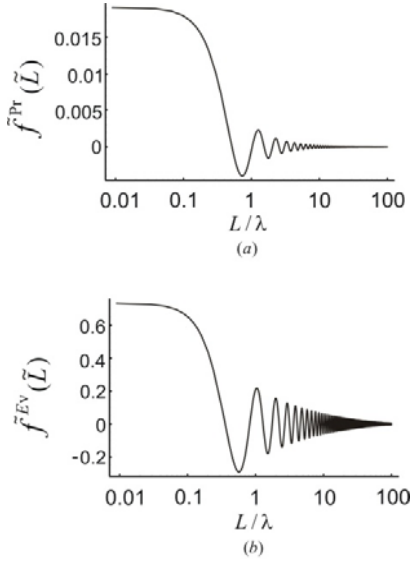

**Fig.3.** Normalised values $\tilde{f}^{Pr}(L)$ - (a) and $\tilde{f}^{Ev}(L)$ - (b) in case of the system out of equilibrium as the functions of normalized lateral separations $L/\lambda$ in accordance with Eqs.(1)-(3).

It can be seen from comparing of figures 2 and 3 that the total traces in equilibrium and out of equilibrium problems are different despite of identical evanescent terms. These total normalized traces $\tilde{W}^{eq}(L) = W^{eq}(L)/W^{eq}(0)$ and $\tilde{W}^{neq}(L) = W^{neq}(L)/W^{neq}(0)$ versus $L/\lambda$ are demonstrated in Fig.4 for equilibrium-(a), (b) and out of equilibrium- (c), (d) problems with use of Eqs.(4) and (1), correspondingly in the case of metal sample. Herewith, the distance dependent terms were computed at $h/\lambda = 0.3$.

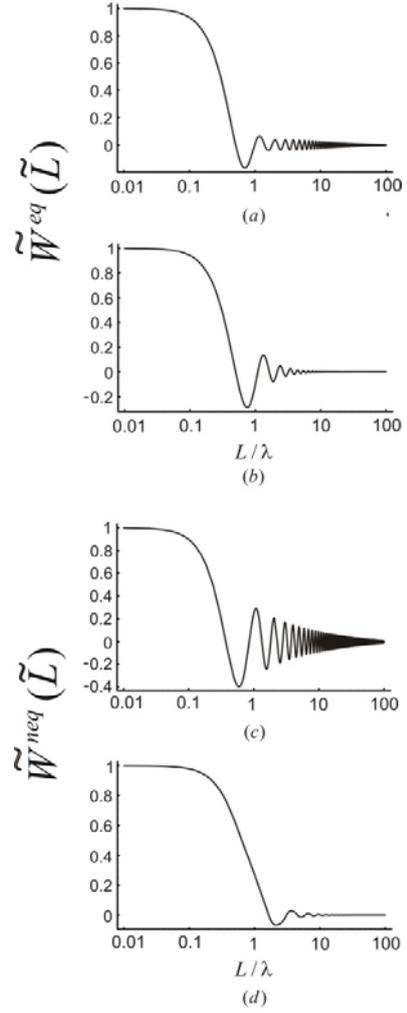

Fig.4. Total normalized traces $\tilde{W}^{eq}(\tilde{L})$ and $\tilde{W}^{neq}(\tilde{L})$ versus $\tilde{L}$ for equilibrium-(a), (b) and out of equilibrium- (c), (d) problems calculated with use of Eqs.(4) and (1), correspondingly in the case of metal sample.

Figures 4a) and 4c) are calculated at the frequency $40000\, cm^{-1}$ relating to the spectral range of surface plasmon excitations ($\text{Re}\{\varepsilon(\omega)\} < -1$). Figures 4b) and 4d) are obtained at the frequency $100000\, cm^{-1}$ which is out of the spectral range of surface excitations. These examples show evidently that correlation functions are different in equilibrium and out of equilibrium problems. Corresponding correlation lengths are anticipated to be different due to obvious difference in the correlation functions. For the system in equilibrium using Eqs.(4) and (8) the relative correlation length along the $L$ direction can be formally represented as follows

$$\ell_{corr}^L = \frac{2\int_0^\infty dL \left| g^{Free}(L) + g^{Pr}(L) + g^{Ev}(L) \right|^2}{\left| g^{Free}(0) + g^{Pr}(0) + g^{Ev}(0) \right|^2}. \quad (10)$$

For the system out of equilibrium we use Eqs.(1) and (8) to obtain the correlation length in this case

$$\ell_{corr}^L = \frac{2\int_0^\infty dL \left| f^{Pr}(L) + f^{Ev}(L) \right|^2}{\left| f^{Pr}(0) + f^{Ev}(0) \right|^2}. \quad (11)$$

Figure 5 exemplifies normalized correlation lengths $\ell_{corr}^L / \lambda$ versus $h/\lambda$ in equilibrium –a) and in out of equilibrium –b) cases calculated with use of Eqs.(10) and (11). In both figures thick curves correspond to the frequency $40000\,cm^{-1}$ and thin curves- to the frequency $100000\,cm^{-1}$.

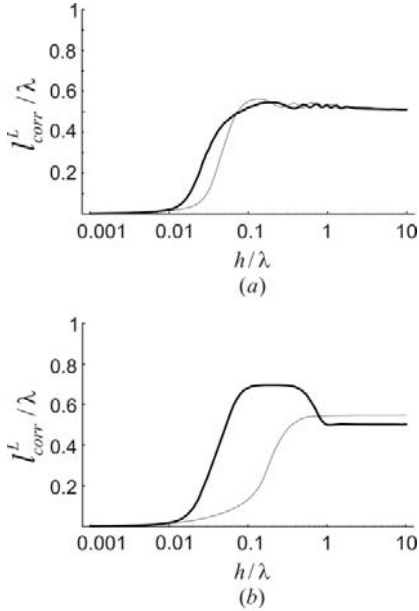

**Fig.5.** Normalized correlation lengths $\ell_{corr}^L / \lambda$ versus $h/\lambda$ in equilibrium –a) and in out of equilibrium system –b) cases calculated with use of Eqs.(10) and (11).

Obviously, the juxtaposition of figures, for instance 5b) with 4c), 4d) does not result in a conclusion on good correspondence between them, especially at the frequency of surface excitations. That is why we make an attempt to modify formulas such as in Eqs. (10) and (11) in order to take an increase of correlations, at least in Fig.4c).

The expressions in Eqs.(1) and (4) for spectral traces mean that corresponding contributions for propagating and evanescent parts are independent. But, the correlation lengths as exprressed by Eq.(10) and (11) allow for their cross-product terms. Hence, we suggest omitting these cross-product terms in Eq.(10) and (11) and using the following expressions in order to take into account the structural independence of each term from all other terms

$$\ell_{corr}^L = \frac{2\int_0^\infty dL \left[ |g^{Free}(L)|^2 + |g^{Pr}(L)|^2 + |g^{Ev}(L)|^2 \right]}{|g^{Free}(0)|^2 + |g^{Pr}(0)|^2 + |g^{Ev}(0)|^2}, \quad (12)$$

$$\ell_{corr}^L = \frac{2\int_0^\infty dL \left[ |f^{Pr}(L)|^2 + |f^{Ev}(L)|^2 \right]}{|f^{Pr}(0)|^2 + |f^{Ev}(0)|^2}. \quad (13)$$

Figure 6 shows normalized correlation lengths $\ell_{corr}^L / \lambda$ versus $h/\lambda$ in equilibrium –a) and in out of equilibrium –b) cases calculated with use of Eqs.(12) and (13). In both figures thick curves correspond to the frequency $40000\,cm^{-1}$ and thin curves- to the frequency $100000\,cm^{-1}$. It is seen that these curves are better matched to correlation functions from Fig.5.

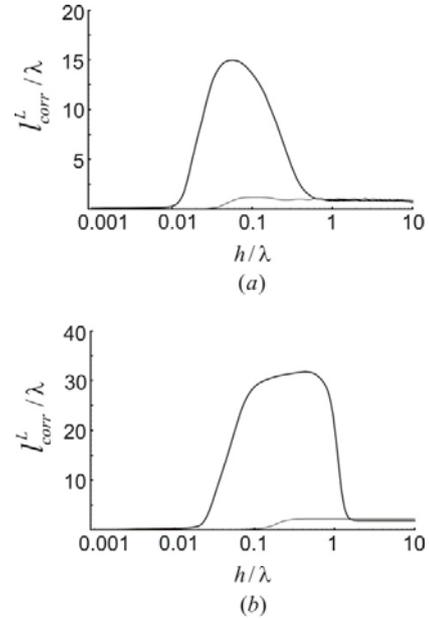

**Fig.6.** Normalized correlation lengths $\ell_{corr}^L / \lambda$ versus $h/\lambda$ in equilibrium –a) and in out of equilibrium system –b) cases in accordance with Eqs.(12) and (13).

We would like especially to emphasize that the known increase of correlations of thermal fields at frequencies of surface polaritons, see figures 4a) and 4c), is more adequately described by the modified Eqs.(12) and (13) as it is exemplified in figure 6.

### 4. Conclusions

Spatial coherence of thermally stimulated fields in far- and near-field zones generated by heated half-space radiating into vacuum is studied in equilibrium and out of equilibrium conditions. It is demonstrated that correlation lengths of fields are different at different thermodynamical conditions. Correlation lengths of thermal fields in vacuum are larger in the system out of equilibrium comparing with the system in equilibrium. We illustrated that in wide range of distances from sample surface the correlation functions are should be calculated using a total sum of evanescent and propagating contributions due to their mutual compensation because of anticorrelations. We modified the often-used formula for correlation length in order to explain an increase of correlation lengths of fields at frequencies of surface polaritons known from a literature. It is demonstrated that correlation lengths as calculated with a proposed formula are in agreement with a behavior of correlation functions of thermal fields in spectral range of surface excitations.